# DISTRIBUTION OF SIZE PARTICLES IN THE GIBBS SYSTEM

V. V. Ryazanov

Institute of Nuclear Research of the National Academy of Sciences of Ukraine,
03680, Kyiv, prosp. Science, 47, Ukraine
E-mail: vryazan@yandex.ru

In the framework of the Gibbs statistical theory, the question of the size of the particles forming the statistical system is investigated. This task is relevant for a wide variety of applications. The distribution for particle sizes and the moments of this quantity are determined from probabilistic considerations. The results are compared with different models and approximations. The particle size depends on the interactions in the system, on the compressibility factor, on the number of interacting particles. The cases of the absence and presence of a rigid incompressible particle core are considered.

*Ключевые слова:* большой канонический ансамбль, гиббсовская статистическая физика, распределение размеров частиц системы, давление, взаимодействия.

1. Введение

Существенным параметром статистической модели служит размер частиц (например, [1-3]). В многочисленных задачах спектроскопии, физической химии, в моделях конденсированного состояния и т.д. возникают вопросы, связанные с определением конечного размера частиц, образующих исследуемую систему. Так, проблема учета конечного объема частиц в связи с исследованием уравнения состояния ядерной материи обсуждалась в [4].

Эффекты исключенного объема (объема, недоступного для центров частиц из-за наличия собственного конечного объема частиц) оказываются очень важными в исследованиях многокомпонентного адронного газа [5].

Связь теории исключенного объема с уравнениями состояния молекулярных систем обсуждается в [6], где концепция исключения объема или площади (для задач поверхности) называется главной. В основу первых уравнений состояния для флюида легли идеи Больцмана и Ван-дер-Ваальса об исключенном объеме, т.е. объеме, на который надо уменьшить реальный объем системы из-за того, что ее частицы имеют собственный объем.

Собственные объемы частиц играют важную роль в исследовании вязкости дисперсных систем [7]. В [8] отмечено, что небольшие изменения радиуса твердой сердцевины частиц или энергии парного взаимодействия могут приводить к заметным изменениям структуры и плотности жидкости. Для решения вопроса о причине высокой плотности гидроциклогексана следует обратиться к законам статистической физики.

Знание размеров частиц важно и во многих задачах биофизики. Отличительная особенность живых систем состоит в том, что биохимические процессы протекают в среде, содержащей высокие концентрации макромолекул (50–400 мг/мл) [9]. При этом важно подчеркнуть, что концентрации макромолекул одного сорта относительно невелики. Однако в совокупности макромолекулы, присутствующие в клетке, занимают значительную часть объема среды (до 40%), вследствие чего доступный объем в клетке сокращается. В английской литературе подобные условия в клетке обозначаются как макромолекулярный краудинг (crowding – to crowd – собираться толпой, набиваться битком). Рассмотрена концепция исключенного объема, позволяющая количественно оценить эффекты краудинга на биохимические реакции.

«Макромолекулярный краудинг» более точно означает «эффект исключенного объема», потому что главная его характеристика – это взаимная непроницаемость всех молекул растворенных веществ. Это неспецифическое стерическое отталкивание присутствует всегда, независимо от других взаимодействий, таких как притяжение или отталкивание, которые возможны между молекулами в растворе. Таким образом, краудинг подобно гравитации нельзя исключить и организмы должны справляться с его последствиями. Влияние исключенного объема на гидродинамические и термодинамические свойства растворов компактных глобулярных белков можно объяснить с помощью моделей, рассматривающих молекулы белка как жесткие частицы сферической формы.

Квазихимическая теория [10] также учитывает конечность объема частиц. Распределение занятого объема для системы твердых сфер получено в [11]. Можно найти сходство и отличие между результатами этой работы и настоящей работой.

Подходы к учету эффектов исключенного объема были предложены в работах [12-15]. Авторами ставилась цель моделирования поведения основных термодинамических величин при наличии у частиц собственных размеров. В работе [16] рассматривались эффекты конечного размера частиц в приближении среднего поля. В [17] рассматривалась задача о плотности вероятности распределения ближайших соседей. В [18] найдено геометрическое условие для предела плотности упаковки в неупорядоченных ансамблях одинаковых жестких выпуклых частиц. В [19] построены аналитические представления модуля всестороннего сжатия, уравнения состояния и удельного логарифма конфигурационного интеграла с учетом наличия у классических частиц эффективного размера (обусловленного отталкиванием на близких расстояниях).

В настоящей работе проблема собственных размеров частиц решается в общем виде для произвольных гиббсовских систем. Размер частиц связывается с термодинамическими характеристиками системы.

2. Функция распределения размеров частиц

Плотность вероятности того, что в формализме большого канонического ансамбля Гиббса в конфигурационном пространстве статистической системы объемом $V$ содержится $n$ частиц с центрами тяжести в точках $r_1,..., r_n$, равна (например, [1])

$$D_n(r_1,...,r_n;V;z) = \frac{z_1 \times ... \times z_n \exp\{-U_n/k_B T\}}{n! Q(z, V, T)}, \qquad (1)$$

где $z_k \sim exp\{\mu_k/k_BT\}$ - активность частицы в $k$-й точке, $\mu_k$ - химический потенциал этой частицы, $k_B$ - постоянная Больцмана, $T$ - абсолютная температура, $U_n$ - потенциальная энергия системы из $n$ частиц. В отсутствии внешнего поля и в приближении парно-аддитивного взаимодействия

$$U_n = \sum_{i \neq j=1}^{n} \varphi_{ij}/2; \qquad \varphi_{ij} = \varphi(|r_i - r_j|); \qquad y_i = \exp\{-\varphi_{ir}/k_BT\}; \qquad y_{ij} = \exp\{-\varphi_{ij}/k_BT\}, \quad (2)$$

где $\varphi_{ij}$ - потенциал парного взаимодействия между частицами в точках $r_i$ и $r_j$, $Q(z;V,T)$ - большая статистическая сумма. Присутствие внешнего поля $U_{ext}(r)$ легко учитывается заменой химического потенциала $\mu_k$ на $\mu_k+U_{ext}(r_k)$ в $z_k$.

$$Q(z;V,T) = 1 + \int_V z_1 dr_1 + \int_V\int_V z_1 z_2 y_{12} dr_1 dr_2 + ... = \exp\{P(z)V/k_BT\}, \quad (3)$$

где $P(z)$ - давление в системе с активностью $z$. Параметрами большого канонического ансамбля, как видно из (1)-(3), являются химический потенциал $\mu_k$ и потенциальная энергия взаимодействия $\varphi_{ij}$ или связанные с ними величины $z_k$ и $y_{ij}$; $V$ - объем системы.

Хотя во многих задачах статистической механики частицы считаются точечными, они обладают конечным объемом, что явно учитывается, например, в уравнении Ван-дер-Ваальса. В статистической механике это учитывается аналитическим видом функции φ(r), ее стремлением к бесконечности при $r \to 0$, введением твердой сердцевины и т. д. [1-3]. Диаметром молекулы принято считать минимальное расстояние, на которое им позволяют сблизиться силы отталкивания. В настоящей работе приводятся вероятностные соображения, позволяющие связать параметры гиббсовского распределения с собственным или эффективным объемом составляющих систему частиц.

Известны выражения для корреляционных функций гиббсовской системы [1], [20]

$$\rho^{(n)}(r_1,...,r_n;V;z) = \prod_{k=1}^{n} z_k e^{-\frac{U_n}{k_BT}} \frac{Q(z\prod_{k=1}^{n} y_k;V)}{Q(z,V)}; \quad (4)$$

$\rho^{(n)}(r_1,...,r_n;V;z)dr_1...dr_n$ является вероятностью того, что при наблюдении за системой, состоящей из $N$ молекул, мы найдем одну молекулу (не обязательно молекулу $1$) в элементе объема $dr_1$ около $r_1$, другую молекулу в элементе объема $dr_2$ около $r_2$,...и, наконец, последнюю молекулу в элементе объема $dr_n$ около $r_n$ [1]. Введем функции $\rho^{(n)}(r_1,...,r_n;\Delta_1,...,\Delta_n)$, представляющие собой совместную плотность вероятности того, что в системе находится $n$ частиц объемом $\Delta_1,...,\Delta_n$ с центрами в точках $r_1,...,r_n$. Предположив независимость событий расположения центров частиц в точках $r_1,...,r_n$ и того, что частица с центром в точке $r_k$ имеет объем $\Delta_k$, запишем соотношение

$$\rho^{(n)}(r_1,...,r_n;\Delta_1,...,\Delta_n;V;z) \sim \rho^{(n)}(r_1,...,r_n;V;z)P^{(n)}(\Delta_1,...,\Delta_n). \qquad (5)$$

где $P^{(n)}(\Delta_1,...,\Delta_n)$ - функция распределения размеров частиц в связанной группе из $n$ частиц.

Событие, состоящее в том, что в системе объемом $V$ находится частица, можно представить, как сумму двух событий: первое - в системе находится частица с эффективным объемом $\Delta(r_1)$ с центром тяжести в точке $r_1$, второе - в системе объемом $V-\Delta(r_1)$ нет частиц. Т.е. одну и ту же частицу (или группу частиц) можно считать как принадлежащей исследуемой системе, так и внешней по отношению к системе без этой группы частиц. Эти два события можно считать независимыми. Но в системе объемом $V-\Delta(r_1)$ присутствует внешнее поле системы объема $\Delta(r_1)$, которое создает находящаяся там частица. Такие рассуждения приводят к соотношениям

$$D_1(r_1;V;z) = \rho^{(1)}(r_1;\Delta_1)D_0(V-\Delta_1;zy_1); \qquad (6)$$

$$D_2(r_1,r_2;V;z) = 2^{-1}\rho^{(1)}(r_1;\Delta_1)D_1(r_2;V-\Delta_1;zy_1) = 2^{-1}\rho^{(2)}(r_1,r_2;\Delta_1,\Delta_2)D_0(V-\Delta_1-\Delta_2;zy_1y_2).$$

Таким же образом для группы (кластера) из $n$ взаимодействующих частиц можно связать вероятность того, что в системе объемом $V$ находится $n$ частиц, и вероятность того, что в системе объемом $V-\Delta_1-...-\Delta_m$ находится $n-m$ ($n \geq m$) частиц, где $\Delta_k$ - эффективный объем одной частицы, соотношениями

$$D_n(r_1,...,r_n;V;z) = n^{-1}\rho^{(1)}(r_1;\Delta_1)D_{n-1}(r_2,...,r_n;V-\Delta_1;zy_1) = [n(n-1)]^{-1}\rho^{(2)}(r_1,r_2;\Delta_1,\Delta_2) \times$$

$$D_{n-2}(r_3,...,r_n;V-\Delta_1-\Delta_2;zy_1y_2) = ... = (n!)^{-1}\rho^{(n)}(r_1,...r_n;\Delta_1,...,\Delta_n)D_0(V-\sum_{k=1}^{n}\Delta_k; z\prod_{k=1}^{n}y_k). \quad (7)$$

Влияние частиц, исключающихся из системы, учитывается как внешнее поле, добавки к химическому потенциалу. Подчеркнем, что это не приближение «самосогласованного поля», а точный учет взаимодействий в соответствующим образом выбранных подсистемах. Множители $(n)^{-1}$, $[n(n-1)]^{-1}$,...,$(n!)^{-1}$ учитывают неразличимость частиц, то, что удаляться может любая из $n$, $n-1$,... частиц (в $D_{n-1}$, например, может отсутствовать любая из $n$ частиц, не обязательно $n$-я).

Подставляя выражения (1) в соотношения (6), (7) получим (из (1): $D_0(V;z,T) = 1/Q(z;V,T)$):

$$\rho^{(n)}(r_1,...,r_n;\Delta_1,...,\Delta_n;V;z) = \prod_{k=1}^{n} z_k e^{-\frac{U_n}{k_B T}} \frac{Q(z\prod_{k=1}^{n} y_k; V - \sum_{k=1}^{n}\Delta_k)}{Q(z;V)} \quad . \tag{8}$$

Подставляя в выражение (8) соотношение (5) и учитывая выражение (4), получаем, что функция распределения размеров частиц в связанной группе из *n* частиц $P^{(n)}(\Delta_1,...,\Delta_n)$ пропорциональна статистической сумме с измененными активностями и объемом

$$P^{(n)}(\Delta_1,...,\Delta_n) \sim Q(z\prod_{k=1}^{n} y_k; V - \sum_{k=1}^{n}\Delta_k) = \exp\{\frac{P(z\prod_{k=1}^{n} y_k)[V - \sum_{k=1}^{n}\Delta_k]}{k_B T}\}, \tag{9}$$

где $P(Az) = k_B T \ln Q(Az;V;T)/V$ - давление при активности *Az*.

Отметим, что соотношения (8) и (1) можно получить функциональным дифференцированием производящего функционала $F(s)=Q(sz;V)[Q(z;V)]^{-1}$, как в [20], если проводить его не по точечным значениям $s(r_k)$, а по объемным $s(\Delta_k)$, и считать, что оно осуществляется по правилу

$$\frac{\delta^n F(s)}{\delta s(\Delta_1)...\delta s(\Delta_n)} = \prod_{k=1}^{n} z_k e^{-\frac{U_n}{k_B T}} \frac{Q(sz\prod_{k=1}^{n} y_k; V - \sum_{k=1}^{n}\Delta_k)}{Q(z,V)}.$$

Тогда

$$\rho^{(n)}(r_1,...,r_n;\Delta_1,...,\Delta_n;V;z) = \frac{\delta^n F(s)}{\delta s(\Delta_1)...\delta s(\Delta_n)}\Big|_{s=1}; \quad n! D_n(r_1,...,r_n;V;z) = \frac{\delta^n F(s)}{\delta s(\Delta_1)...\delta s(\Delta_n)}\Big|_{s=0},$$

как в [20].

Плотность вероятности функции распределения $P^{(n)}(\Delta_1,...,\Delta_n)$ определяется из (9)

$$p^{(n)}(\Delta_1,...,\Delta_n) = \frac{1}{\alpha^{(n)}} \frac{\partial^n Q(z\prod_{k=1}^{n} y_k; V - \sum_{k=1}^{n}\Delta_k)}{\partial \Delta_1...\partial \Delta_n}, \tag{10}$$

где

$$\alpha^{(n)} = \int...\int \frac{\partial^n Q(z\prod_{k=1}^{n} y_k; V - \sum_{k=1}^{n}\Delta_k)}{\partial \Delta_1...\partial \Delta_n} d\Delta_1...d\Delta_n,$$

- нормировочный множитель. Интегрирование в $\alpha^{(n)}$ должно проводиться с учетом того, что величины $\Delta_1,...,\Delta_n$ занимают определенный объем, т.е.

$$\int ... \int (...) d\Delta_1 ... d\Delta_n = \int_0^V d\Delta_1 \int_0^{V-\Delta_1} d\Delta_2 ... \int_0^{V-\Delta_1-...-\Delta_{n-1}} (...) d\Delta_n . \qquad (11)$$

При этом

$$\frac{\partial^n Q(z\prod_{k=1}^{n} y_k; V - \sum_{k=1}^{n}\Delta_k)}{\partial \Delta_1 ... \partial \Delta_n} = (-\alpha_n)^n e^{\alpha_n(V-\Delta_1-...-\Delta_n)}; \qquad \alpha_n = P(zy_1...y_n)/k_B T ;$$

$$\alpha^{(n)} = (-1)^{n-1} [\sum_{k=0}^{n-1} \frac{V^k \alpha_n^k}{k!} - \exp\{\alpha_n V\}].$$

Тогда из (9)-(10) получаем, что плотность вероятности

$$p^{(n)}(\Delta_1,...,\Delta_n) = \frac{(-\alpha_n)^n \exp\{\alpha_n(V-\Delta_1-...-\Delta_n)\}}{\alpha^{(n)}} .$$

При *n=1*:

$$p^{(1)}(\Delta_1) = \frac{\alpha_1 \exp\{-\alpha_1 \Delta_1\}}{1 - \exp\{-\alpha_1 V\}};$$

$$n=2: \qquad p^{(2)}(\Delta_1,\Delta_2) = \frac{\alpha_2^2 \exp\{-\alpha_2(\Delta_1+\Delta_2)\}}{1 - \exp\{-\alpha_2 V\}(1+\alpha_2 V)};$$

$$n=3: \qquad p^{(3)}(\Delta_1,\Delta_2,\Delta_3) = \frac{\alpha_3^3 \exp\{-\alpha_3(\Delta_1+\Delta_2+\Delta_3)\}}{1 - \exp\{-\alpha_3 V\}(1+\alpha_3 V + (\alpha_3 V)^2/2)};...;$$

$$p^{(r)}(\Delta_1,...,\Delta_r) = \frac{\alpha_r^r \exp\{-\alpha_r(\Delta_1+\Delta_2+...+\Delta_r)\}}{1 - \exp\{-\alpha_r V\}\sum_{k=0}^{r-1}\frac{(\alpha_r V)^k}{k!}}; \; r=1,2,... . \qquad (12)$$

Используя формулу для разложения Тейлора [14]:

$$e^{\alpha_r V} = \sum_{k=0}^{r}\frac{(\alpha_r V)^k}{k!} + R_r ; \qquad R_r(\alpha_r V) = \frac{(\alpha_r V)^{r+1}}{(r+1)!} e^{\theta_r \alpha_r V}; \quad 0 < \theta_r < 1 ,$$

где $R_r(\alpha_r V)$ - остаточный член ряда Маклорена (форма Лагранжа остаточного члена), получаем:

$$\alpha^{(n)} = (-1)^n R_{n-1}(\alpha_n V);$$

$$p^{(r)}(\Delta_1,...,\Delta_r) = \frac{\alpha_r^r}{R_{r-1}(\alpha_r V)} e^{\alpha_r(V-\Delta_1-...-\Delta_r)} = \frac{r!}{V^r} e^{\alpha_r[V(1-\theta_{r-1})-\Delta_1-...-\Delta_r]}. \qquad (13)$$

Для формы остаточного члена [21]

$$R_n(x) = \int_0^x \frac{e^x}{n!}(x-t)^n dt = \gamma(n+1,x)\frac{e^x}{n!},$$

где $\gamma(\alpha,x)$ - неполная гамма-функция [21], $\gamma(n+1,x) = n!(1 - e^{-x}\sum_{k=0}^{n}\frac{x^k}{k!})$, получаем, что

$$p^{(r)}(\Delta_1,...,\Delta_r) = \frac{\alpha_r^r(r-1)!}{\gamma(r,\alpha_r V)}\exp\{-\alpha_r(\Delta_1 + ... + \Delta_r)\}. \qquad (14)$$

Так как $n$ – целое число, то [21]

$$R_n(x) = e^x(1 - e^{-x}\sum_{k=0}^{n}\frac{x^k}{k!}).$$

Подстановка этого выражения в соотношение (13) приводит к ранее полученной формуле (12). Заметим, что выражение (14) может описывать комплексы частиц и в случае дробных значений их числа, величины $r$.

### 3. Среднее значение и дисперсия объема частицы при отсутствии жесткой несжимаемой сердцевины частиц

Если определить при помощи выражений (10)-(14) среднее значение объема частицы $\Delta_1$ в группе из $n$ связанных частиц

$$\overline{\Delta_1^{(n)}} = \int...\int \Delta_1 p^{(n)}(\Delta_1,...,\Delta_n)d\Delta_1...d\Delta_n,$$

то получим (интегрирование проводится по правилу (11))

$$\overline{\Delta_1^{(n)}} = \frac{1}{\alpha_n}\frac{R_n(\alpha_n V)}{R_{n-1}(\alpha_n V)}. \qquad (15)$$

При $n=1$

$$\overline{\Delta_1^{(1)}} = \frac{[1 - (1+\alpha_1 V)\exp\{-\alpha_1 V\}]}{\alpha_1(1 - \exp\{-\alpha_1 V\})}. \qquad (16)$$

При объеме системы $V\to\infty$, $\overline{\Delta_1^{(1)}} = 1/\alpha_1$; для идеального газа $\overline{\Delta_1^{(1)}} = v = V/N = 1/\rho$, где $N$ - полное число частиц в системе, $\rho$ - средняя плотность; при $V\to 0$, $\overline{\Delta_1^{(1)}} \to 0$, что согласуется с физическими представлениями. Так же и $\overline{\Delta_1^{(n)}} = 1/\alpha_n$ при $V\to\infty$; для идеального газа $\overline{\Delta_1^{(n)}} = v = V/N = 1/\rho$. При $V\to 0$ $\overline{\Delta_1^{(n)}} \to 0$. Подставляя в формулу (15) явные выражения для $R_n$, $R_{n-1}$, получим

$$\overline{\Delta_1^{(n)}} = \frac{1}{\alpha_n}\frac{1 - \exp\{-\alpha_n V\}\sum_{k=0}^{n}\frac{(\alpha_n V)^k}{k!}}{1 - \exp\{-\alpha_n V\}\sum_{k=0}^{n-1}\frac{(\alpha_n V)^k}{k!}} = \frac{1}{\alpha_n}[1 - \frac{\exp\{-\alpha_n V\}\frac{(\alpha_n V)^n}{n!}}{1 - \exp\{-\alpha_n V\}\sum_{k=0}^{n-1}\frac{(\alpha_n V)^k}{k!}}]. \qquad (17)$$

При $n\to\infty$, $\overline{\Delta_1^{(n)}}\to 0$. Если $n\to\infty$ и $V\to\infty$, то $\overline{\Delta_1^{(n)}}\to V/n$. При $\alpha_n\to 0$, $\overline{\Delta_1^{(n)}}\to V/n$; $\alpha_n\to\infty$, $\overline{\Delta_1^{(n)}}\to 0$. При $n=1$:

$$\overline{\Delta_1^{(1)}} = \frac{1}{P/k_BT}\frac{(e^{PV/k_BT}-1-PV/k_BT)}{(e^{PV/k_BT}-1)} \approx \frac{k_BT}{P}; \quad P = P(zy_1).$$

Для других значений $n$ получаем, используя выражение для остаточного члена ряда Тейлора в форме Лагранжа,

$$\overline{\Delta_1^{(n)}} = \frac{V}{(n+1)}e^{-\frac{P(zy_1...y_n)V(\theta_{n-1}-\theta_n)}{k_BT}} = \frac{V}{(n+1)}e^{-\alpha_n V(\theta_{n-1}-\theta_n)},$$

где $\theta_{n-1}>\theta_n$; $0<\theta_{n-1}<1$; $0<\theta_n<1$.

Дифференцируя выражение (15) по $\alpha_n$, получим, что

$$\frac{\partial \alpha_n}{\partial V} = \frac{\alpha_n}{\frac{1}{f_R}(\overline{\Delta_1^{(n)}} + \alpha_n\frac{\partial\overline{\Delta_1^{(n)}}}{\partial\alpha_n})-V}; \quad f_R = \frac{\partial(R_n(x)/R_{n-1}(x))}{\partial x} = \frac{x^{n-1}}{(n-1)!}\frac{[(x/n)R_{n-1}(x)-R_n(x)]}{R^2_{n-1}(x)}\bigg|_{x=\alpha_n V}.$$

Фазовый переход происходит при $\frac{\partial P}{\partial v}\leq 0; v=\frac{V}{\overline{N}}$. Если среднее полное число частиц в системе $\overline{N}$ постоянно, то $\frac{\partial \alpha_n}{\partial V}\sim\frac{\partial P}{\partial v}$. Величина $\alpha_n>0$, и фазовый переход наступает при

$$\frac{1}{f_R}(\overline{\Delta_1^{(n)}} + \alpha_n\frac{\partial\overline{\Delta_1^{(n)}}}{\partial\alpha_n})-V\leq 0.$$

Величину $\alpha_n = P(zy_1...y_n)/k_BT$ можно связать с корреляционными функциями $\rho^{(n)}$. Из выражений (4), (5), (8), если в соотношении (5) стоит знак равенства, вытекает, что

$$P^{(n)}(\Delta_1,...,\Delta_n) = \frac{Q(zy_1...y_n, V-\Delta_1-...-\Delta_n)}{Q(zy_1...y_n, V)} = e^{-\alpha_n(\Delta_1+...+\Delta_n)}.$$

Подставляя это выражение в (4), (5), (8), получаем, что (этот результат сразу следует из (3), (4); соотношения, связывающие $\alpha_n$ с $\alpha_0$, получены также в [1])

$$\rho^{(n)}(r_1,...,r_n;V;z) = \prod_{k=1}^{n}z_k e^{-U_n/k_BT+(\alpha_n-\alpha_0)V}; \quad \alpha_n = \alpha_0+[\ln(\rho^{(n)}/z_1...z_n)+U_n/k_BT]/V; \quad \alpha_0 = P(z)/k_BT;$$

$$\overline{\Delta_1^{(n)}} = \frac{V}{n+1}e^{-[PV/k_BT+\ln(\rho^{(n)}/z_1...z_n)+U_n/k_BT](\theta_{n-1}-\theta_n)}.$$

Из (17) видно, что при разных $n$ зависимость собственного объема частиц от давления различна. Выражение для $\overline{\Delta_1^{(n)}}$ записывается также и в другом виде. Явная зависимость $\theta_n$ и $\theta_{n-1}$ от термодинамических параметров не определяется из вида остаточного члена.

Поэтому выражение остаточного члена через неполную гамма-функцию может быть предпочтительней.

Из (10)-(14) определяется также дисперсия значения $\Delta_1^{(n)}$

$$D^{(n)}_\Delta = \overline{\Delta^{(n)2}_1} - (\overline{\Delta^{(n)}_1})^2 = \frac{1}{\alpha_n^2}\left[\frac{2R_{n+1}(\alpha_n V)}{R_{n-1}(\alpha_n V)} - \left(\frac{R_n(\alpha_n V)}{R_{n-1}(\alpha_n V)}\right)^2\right].$$

При $n=1$: $D_\Delta^{(1)} = \overline{\Delta^{(1)2}_1} - (\overline{\Delta^{(1)}_1})^2 = \dfrac{e^{2\alpha_1 V} + 1 - (\alpha_1 V)^2 e^{\alpha_1 V} - 2e^{\alpha_1 V}}{\alpha_1^2(e^{2\alpha_1 V} + 1 - 2e^{\alpha_1 V})} \approx \dfrac{1}{\alpha_1^2}$.

Для величины

$$\overline{\Delta^{(n)2}_1} = \frac{2}{\alpha_n^2}\frac{R_{n+1}[\alpha_n V]}{R_{n-1}[\alpha_n V]}$$

получаем при больших значениях $\alpha_n V$

$$\overline{\Delta^{(n)2}_1} \sim \frac{2}{\alpha_n^2} ; \qquad D^{(n)}_\Delta \sim \frac{1}{\alpha_n^2} ;$$

и при произвольных $\alpha_n V$:

$$\overline{\Delta^{(n)2}_1} = \frac{2}{\alpha_n^2}\left(1 - \frac{e^{-x}\dfrac{x^n}{n!}\left(1+\dfrac{x}{n+1}\right)}{1 - e^{-x}\left(1 + x + \ldots + \dfrac{x^{n+1}}{(n+1)!}\right)}\right)\Bigg|_{x=\alpha_n V} = \frac{2V^2}{(n+2)(n+1)}e^{-\alpha_n V(\theta_{n-1} - \theta_{n+1})} ;$$

Из (10)-(14) записывается соотношение $\overline{\Delta^{(n)}_1 + \ldots + \Delta^{(n)}_n} = \overline{\Delta^{(n)}_1} + \ldots + \overline{\Delta^{(n)}_n}$, корреляторы вида $\left\langle \Delta^{(n)}_1 \Delta^{(n)}_2 \right\rangle$, а также другие соотношения и функции от $\Delta_1, \ldots, \Delta_n$.

Кроме выражений вида (17) можно получить другие выражения и для распределения размера частиц, отличные от (12), и для среднего размера. Так, из выражений (4)-(9) видно, что распределение размеров частиц представляет собой многомерное, определенное на конечном интервале (урезанное) показательное распределение. Если определить нормировку этого распределения, как указано в [22], а не как в соотношениях (10)-(11), то плотность вероятности распределения размеров частиц будет иметь вид

$$p^{(n)}(\Delta_1, \ldots, \Delta_n) = \frac{\alpha_n^n \exp\{-\alpha_n(\Delta_1 + \ldots + \Delta_n)\}}{(1 - \exp\{-\alpha_n V\})^n} .$$

Среднее значение

$$\overline{\Delta}_1^{(n)} = k_n / \alpha_n, \quad k_n = [1 - (1 + \alpha_n V)\exp\{-\alpha_n V\}] / (1 - \exp\{-\alpha_n V\}) .$$

Сравнивая это выражение с выражением (17), найдем, что они различаются множителем $k_n$, в (17) $k_n = \dfrac{1-\exp\{-\alpha_n V\}\sum\limits_{k=0}^{n}\dfrac{(\alpha_n V)^k}{k!}}{1-\exp\{-\alpha_n V\}\sum\limits_{k=0}^{n-1}\dfrac{(\alpha_n V)^k}{k!}}$.

$$\alpha_n = \dfrac{P(z)}{kT} + \dfrac{1}{V}\ln\dfrac{\rho^{(n)}}{z^n \exp\{-U_n/kT\}} = \dfrac{P(z)}{kT} + C_n, \quad C_n = -\dfrac{U_n}{VkT} + \dfrac{1}{V}\ln\dfrac{\rho^{(n)}}{z^n}.$$

### 4. Моменты размеров частиц с жесткой несжимаемой сердцевиной

Рассмотрим теперь случай, когда у частиц присутствует жесткая несжимаемая сердцевина объемом $d$ (т.е. $\Delta \geq d$; выше полагалось $d=0$), то интегрирование проводится в пределах $\int\limits_{d}^{V-d(n-1)}d\Delta_1 \int\limits_{d}^{V-d(n-2)-\Delta_1}d\Delta_2 ... \int\limits_{d}^{V-\Delta_1-...-\Delta_{n-1}}(...)d\Delta_n$. Тогда

$$p_d^{(r)}(\Delta_1,...,\Delta_r) = \dfrac{\alpha_r^r}{R_{r-1}[\alpha_r(V-rd)]}e^{\alpha_r(V-\Delta_1-...-\Delta_r)};$$

$$\overline{\Delta_1^{(1)}}_d = \int_d^V \Delta_1 p_d^{(1)}(\Delta_1)d\Delta_1 = \dfrac{[e^{-d\alpha_1}-(1+\alpha_1(V-d))e^{-V\alpha_1}]}{\alpha_1[e^{-d\alpha_1}-e^{-V\alpha_1}]} \approx d + \dfrac{1}{\alpha_1}. \quad (18)$$

Для идеального газа $\rho=z$; $\alpha_1 = P/k_B T = \alpha_0 = 1/(\overline{\Delta_1^{(1)}}_d - d) = \rho$; $\overline{\Delta_1^{(1)}}_d = d + 1/\rho$. При произвольных $n$:

$$\overline{\Delta_1^{(n)}}_d = \dfrac{1}{\alpha_n}(\dfrac{R_n[\alpha_n(V-nd)]}{R_{n-1}[\alpha_n(V-nd)]} + d\alpha_n) = d + \overline{\Delta_1^{(n)}}(V-nd) =$$

$$= d + \dfrac{1}{\alpha_n}\dfrac{1-\exp\{-\alpha_n(V-nd)\}\sum\limits_{k=0}^{n}\dfrac{(\alpha_n(V-nd))^k}{k!}}{1-\exp\{-\alpha_n(V-nd)\}\sum\limits_{k=0}^{n-1}\dfrac{(\alpha_n(V-nd))^k}{k!}} = d + \dfrac{(V-nd)}{n+1}e^{-(\theta_{n-1}-\theta_n)\alpha_n(V-nd)}, \quad (19)$$

где $\overline{\Delta_1^{(n)}}(V)$ дано в (16). Соотношения (15)-(16) получаются из (18)-(19) при $d=0$. Можно связать величину $\Delta$ с теорией свободного объема [1] и с решеточными теориями.

Величина $\overline{\Delta^{(n)}_1}_d$ зависит от $n$ - числа частиц в группе взаимодействующих частиц. При этом $\overline{\Delta^{(n)}_1}_d - d \leq 1/\alpha_n$; $\overline{\Delta^{(n)}_1}_d - d = K_n/\alpha_n$; $0 \leq K_n = R_n[\alpha_n(V-nd)]/R_{n-1}[\alpha_n(V-nd)] \leq 1$. При $\alpha_n \to 0$, $\overline{\Delta_1^{(n)}}_d \to V/n$, не зависит от $d$; $\alpha_n \to \infty$, $\overline{\Delta_1^{(n)}}_d \to d$. Максимальное значение $n^{max}=V/d$, т.к. $dn \leq V$. При $n \to n^{max}$, $\alpha_n V(1-nd/V) \to 0$; $\overline{\Delta^{(n)}_1}_d \to d$. Величина $\overline{\Delta^{(n)}_1}_d - d$ в зависимости от $n$ ведет себя, как показано на рис.1а-1б, где обозначено $r=\overline{\Delta^{(n)}_1}_d - d$, $a=\alpha_n$. При расчете использовалась модель упругих шаров [2] с потенциалом вида $\Phi(r)=0$, $r>r_0$; $\Phi(r)=\infty$, $r \leq r_0$.

Тогда можно не учитывать зависимость $\alpha_n$ от $n$. При этом $\partial \overline{\Delta^{(n)}_1}_d/\partial n \leq 0$. На рис.1в показана ситуация, когда при некоторых $n$ $\partial \overline{\Delta^{(n)}_1}_d/\partial n > 0$. Для газов $n \sim 1$, значения $n$ велики для твердого тела, и $n$ занимает промежуточные значения для жидкостей. Можно найти значения $n$, которые соответствовали бы приближениям идеального газа, Ван-дер-Ваальса, аппроксимантам Паде, модели твердых сфер и т.д.

Аналогичная зависимость $\overline{\Delta^{(n)}_1}_d - d$ от $\alpha_n$: при $\alpha_n \to 0$, $\overline{\Delta^{(n)}_1}_d \to V/n$; при $\alpha_n \to \infty$, $\overline{\Delta^{(n)}_1}_d - d \to 0$, т.е. зависимость $\overline{\Delta^{(n)}_1}_d - d$ от $\alpha_n$ ведет себя, как показано на рис.2. Так как $\alpha_n \to 0$ при $\varphi_{12} \to \infty$, а $\alpha_n \to \alpha_{n=0}$ при $\varphi_{12} \to 0$, то так же (но $\overline{\Delta^{(n)}_1}_d - d$ растет с ростом $\varphi_{12}$) ведет себя зависимость $\overline{\Delta^{(n)}_1}_d - d$ от $y_{12}$. Возможны ситуации, при которых величина $\overline{\Delta^{(n)}_1}_d - d$ может расти с ростом $\alpha_n$. Давление ($\sim \alpha_n \sim P/k_B T$) "сжимает" частицу до минимального значения, равного $d$.

## 5. Заключение

Диаметр частиц $r_0$ занимает важное место в определении "размеров" сплошной среды [3]. Например, в приближении $\overline{\Delta^{(n)}_1} \approx V/N$ трудно провести различие между микромасштабными и крупномасштабными флуктуациями [3]. В различных приложениях результаты настоящей работы можно использовать для систем, описывающихся распределением Гиббса. Так, в задачах физической химии или для коллоидных частиц – в той мере, в которой для них справедлива гиббсовская статистика. Полученные результаты могут оказаться полезными при исследовании ряда задач физической химии и химической физики. Например, для выяснения механизма химических реакций, определения путей транспорта ионов через мембраны, установления структуры поверхности раздела фаз, сорбции и транспорта молекул в пористых структурах, микроструктуры поверхностных слоев жидкостей, микродинамики процессов гидратации ионов и в других актуальных проблемах знание размеров составляющих систему частиц и зависимости этих размеров от параметров системы должны внести определенный вклад в изучение этих задач.

Таким образом, основные результаты гиббсовской статистической физики (образование кластеров и различные приближения для уравнения состояния, отрицательность зависимости $\partial P/\partial v$, фазовые переходы, теория флуктуаций и т.д.) находят соответствие в эффективном объеме частиц, изучение которого может служить независимым способом определения искомых свойств исследуемой системы. Следует провести более детальное сравнение с результатами работ [1-19], где явно или неявно присутствует размер частиц (например, сопоставление результатов настоящей работы и работы [16] показывает их соответствие), а также более полно использовать известные соотношения гиббсовской статистики для получения явных выражений для $\overline{\Delta^{(n)}}_1$ при заданных моделях уравнений состояния.

Размер частиц зависит от взаимодействий в системе. В настоящей работе рассмотрены гиббсовские распределения и равновесные состояния. Не оценивается возможное влияние неравновесных эффектов на размер частиц. Показательное распределение для размеров частиц получено из гиббсовской статистики. Если же исходить, например, из распределений Цаллиса [23], то распределение для размеров частиц будет иметь степенной характер.

## СПИСОК ЛИТЕРАТУРЫ

**Рисунки** к статье В.В. Рязанова «Распределение размеров частиц в гиббсовской системе».

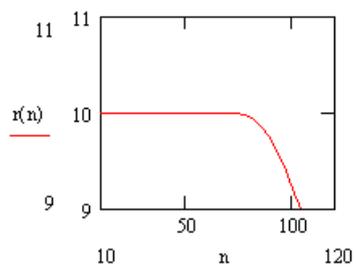

Рис. 1а. n=10,…,120; a=$10^{-1}$ ; d=$10^{-5}$ ; V=$10^{3}$ .

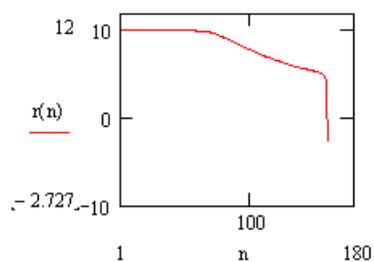

Рис. 1б. n=1,2,…,180; a=$10^{-1}$ ; d=$10^{-2}$ ; V=$10^{2.91}$ .

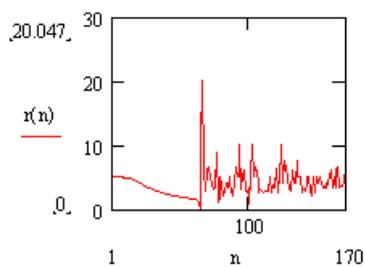

Рис.1в. n=1,2,…,170; a=$10^{-0.7}$ ; d=$10^{-3}$; V=$10^{2}$

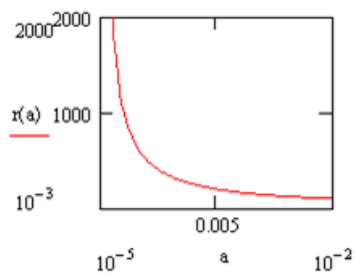

Рис.2а. a=$10^{-5}$, $10^{-4}$, …, $10^{-2}$ ; d=$10^{-7}$; n=10 ; V=$10^{17}$

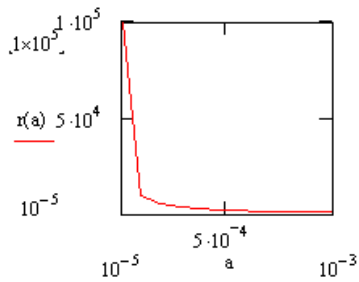

Рис. 2б.  a=$10^{-5}$, $10^{-4}$; …,$10^{-3}$;   d=$10^{-5}$;  V=$10^{10}$;  n=30 .

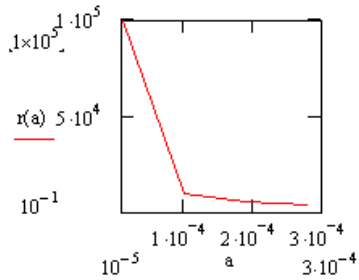

Рис. 2в.  a=$10^{-5}$,…,$3·10^{-4}$;   d=$10^{-1}$;  V=$10^{10}$ ;  n=47 .

**Подписи** к РИСУНКАМ к статье В.В. Рязанова «Распределение размеров частиц в гиббсовской системе».

Рис.1. Зависимость разницы между средним размером частицы $\overline{\Delta^{(n)}}_{1\,d}$ в кластере $n$ взаимодействующих частиц в модели с жесткой несжимаемой сердцевиной объемом $d$ и величиной $d$ от $n$ - числа частиц в группе взаимодействующих частиц. Использована модель взаимодействия упругих шаров [2]; $\partial \overline{\Delta^{(n)}}_{1\,d}/\partial n \leq 0$; r=$\overline{\Delta^{(n)}}_{1\,d}$-d; a=$\alpha_n$=P($zy_1...y_n$)/$k_BT$.

Рис. 1а: $\partial \overline{\Delta^{(n)}}_{1\,d}/\partial n \leq 0$; a=$10^{-1}$; d=$10^{-5}$; V=$10^3$; 10≤n≤120;  Рис. 1б: $\partial \overline{\Delta^{(n)}}_{1\,d}/\partial n \leq 0$; a=$10^{-1}$; d=$10^{-2}$; V=$10^{2.91}$; 1≤n≤180;   Рис. 1в: при некоторых $n$  $\partial \overline{\Delta^{(n)}}_{1\,d}/\partial n > 0$, фазовый переход; a=$10^{-0.7}$; d=$10^{-3}$; V=$10^2$; 1≤n≤170.

Рис.2. Зависимость разницы между средним размером частицы $\overline{\Delta^{(n)}}_{1\,d}$ в кластере $n$ взаимодействующих частиц в модели с жесткой несжимаемой сердцевиной объемом $d$ и величиной $d$ от $a=\alpha_n$ – фактора сжимаемости (умноженного на $R/k_BV$, $R$ – универсальная газовая постоянная); r=$\overline{\Delta^{(n)}}_{1\,d}$-d; a=$\alpha_n$=P($zy_1...y_n$)/$k_BT$.

Рис. 2а: $10^{-5} \leq a \leq 10^{-2}$; d=$10^{-7}$; n=10; V=$10^{17}$. Рис. 2б: $10^{-5} \leq a \leq 10^{-3}$; d=$10^{-5}$; n=30; V=$10^{10}$. Рис. 2в: $10^{-5} \leq a \leq 3·10^{-4}$; d=$10^{-1}$; n=47; V=$10^{10}$.